# Searching for High-Tc Superconductivity in the Z = 8.67 Family

O. P. Isikaku-Ironkwe[1, 2]
[1]The Center for Superconductivity Technologies (TCST)
Department of Physics,
Michael Okpara University of Agriculture, Umudike (MOUAU),
Umuahia, Abia State, Nigeria
and
[2]RTS Technologies, San Diego, CA 92122

## Abstract

$CaLi_2$, a member of the Z =8.67 family, with Ne =1.333 is superconducting at 14K at 40 GPA pressure. Through computational combinatorial synthesis of the Z =4.67 and Z=12.67 families, we create new members of the Z =8.67 family. We discover over 160 potential members of the Z = 8.67 family with Ne=2.67 that have more than three elements. Our estimations suggest that most of them will be superconducting with Tcs between 30K and 60K at ambient pressure.

## Inspiration

*"The continued experimental search for new materials and new behavior in binary, ternary, quaternary and other compounds is one of the most important engines driving progress in condensed matter Physics"*

-----Canfield & Crabtree (Physics Today, March 2003)

## Introduction

The discovery of high-Tc superconductivity at 39K in $MgB_2$ [1] in 2001 was a significant landmark in the continual search for high-temperature superconductivity in low Z materials [2 - 7]. Though efforts have been made to create $MgB_2$-like superconductivity, certain DFT predictions [8 - 11] based on structural and isoelectronic symmetry have not yielded the desired high-Tc results. Experimental searches [12 - 16] based too on structural and isoelectronic symmetry with $MgB_2$ have also failed to find high-Tc superconductivity in low-Z materials. Following a completely new approach, we explored superconductivity from the chemical building blocks of electronegativity, valence electrons and atomic number. This led to the discovery of a far reaching formula for Tc in material specific terms and symmetry rules [17] which we have used in discovering and classifying new families of materials and **superconductors [18 - 27]. In this paper we continue the search for novel high-Tc superconductivity in an under explored family of materials termed Z=8.67. We show herein that by computational combinatorial synthesis of two families of materials: the Z=4.67 and Z=12.67 families, we get the Z=8.67 family. We also show that the Z=8.67 family has binary, ternary and more-than-three elements members whose Tcs may be higher than 60K. This paper is structured as follows: first we review the material specific framework on which our searching for new superconductors is based. Next we show that $CaLi_2$ falls within the Z =8.67 family. We proceed to create new members of this family using the material specific search framework that involves computational material synthesis of Z=4.67 and Z=12.67 families. Finally, we estimate the Tcs of some of the resultant new Z=8.67 family of materials using the Tc equation from the material specific search framework.**

## Material Specific Search Framework

**The Tc of a compound can be estimated [17] from the equation:**

$$T_c = \mathcal{X} \frac{\mathrm{Ne}}{\sqrt{Z}} K_o \qquad (1)$$

**where $\mathcal{X}$ represents average electronegativity, Ne the average valence electron count and Z the average atomic number of the compound. Ko is a parameter with value n(Fw/Z), where Fw is the formula weight of the compound and n is an empirically estimated number. For $MgB_2$, n =3.65. These averages are given [17], for a compound $A_pB_qC_rD_sE_t$, as:**

$$\mathcal{X} = \frac{p\mathcal{X}_A + q\mathcal{X}_B + r\mathcal{X}_C + s\mathcal{X}_D + t\mathcal{X}_E}{p+q+r+s+t} \qquad (2)$$

$$\mathrm{Ne} = \frac{pNe_A + qNe_B + rNe_C + sNe_D + tNe_E}{p+q+r+s+t} \qquad (3)$$

$$Z = \frac{pZ_A + qZ_B + rZ_C + sZ_D + tZ_E}{p+q+r+s+t} \qquad (4)$$

**Here p, q, r, s, t represent the numbers of atoms present in the elements A, B, C, D and E respectively. The formula weight, Fw, is given as:**

$$Fw = pFw_A + qFw_B + rFw_C + sFw_D + tFw_E \qquad (5)$$

The material specific characterization dataset (MSCD) is constructed from the above data and displayed as:

MSCD of compound = $\langle \mathcal{X}$, Ne, Z, Ne/$\sqrt{Z}$ , Fw, Fw/Z$\rangle$ (6)

## Superconductivity in $CaLi_2$

The search for superconductivity in low Z materials revealed that at 40 GPA, lithium calcium alloy, $CaLi_2$, becomes superconducting [28] at 13K. Using the Material Specific Characterization Dataset (MSCD) scheme [17] we find that $CaLi_2$ belongs to the Z =8.67 family where Z is average atomic number. The existence, at high pressure, of superconductivity in $CaLi_2$ with low electronegativity, low valence electrons and low Ne/$\sqrt{Z}$ suggests that higher values of those parameters with Z=8.67 may give higher Tcs [17] at ambient pressure. This inspired our search for Z=8.67 materials using the MSCD above and the computational combinatorial synthesis method described in the next section.

## Computational Combinatorial Synthesis of Z =8.67 Family

We propose that the combinatorial synthesis of two families: Z=4.67 and Z=12.67 materials, with Ne=2.67 will produce a Z=8.67 material according to the equation:

$$F_{Z=4.67} + F_{Z=12.67} = F_{Z=8.67} \quad (7)$$

The MSCD of the resultant $F_{Z=8.67}$ compounds is given by:

MSCD of $F_{Z=8.67}$ compound = $\langle \mathcal{X}$, Ne, Z, Ne/$\sqrt{Z}$ , Fw, Fw/Z$\rangle$

With Ne = 2.667 in both $F_{Z=4.67}$ and $F_{Z=12.67}$ compounds in Tables 1 and 2, we get:

MSCD of $F_{Z=8.67}$ compounds = $\langle \mathcal{X}$, 2.667, 8.667, 0.9058, Fw, Fw/Z$\rangle$ (8)

where Ne/$\sqrt{Z}$ = 0.9058. A high-Tc occurs [17] when:

$$0.8 < Ne/\sqrt{Z} < 1.0 \quad (9)$$

The combinatorial process that yields the MSCDs of the new F=8.67 compounds is shown in Figure 1. The potential products, some of which are shown in Tables 3 and 4, are discussed in the next section.

## Results

The combinatorial synthesis of ten $F_{Z=4.67}$ and sixteen $F_{Z=12.67}$ families of materials yields 160 possible materials with Z=8.67, some of which are shown in Table 3. From the MSCDs of the Z=8.67 materials in Table 4, we can estimate their Tcs using equation (1) above, assuming Ko = 2.5(Fw/Z). The results shown in Table 4 indicate that at the value of n=2.5, the Tcs are in the range 45 – 48K. For n =3.5, the highest Tc =67.9K for $Be_2NaCaNC$. Thus we should expect Tcs between 45 and 68K for this family of materials with Fw/Z of 12.4.

## Discussion

Low Tc materials have long been proposed as possible high-Tc superconductors [2]. However we showed [17, 18] that low Z is not enough to guarantee high-Tc superconductivity. The material must also meet the condition: $0.8<Ne/\sqrt{Z}<1.0$. In this paper, the Z=8.67 materials meet this condition and also have electronegativity of over 1.6. In this study, we have not used the traditional DFT methods to arrive at our empirical estimates. Rather we have used the material specific formula for Tc derived earlier [17]. The thermodynamics of phases have been temporally ignored in obtaining the estimates. However the observed symmetry in the materials MSCDs strongly suggests that they are feasible and thus require further studies. Some of them may require high-pressure-high-temperature synthesis in an argon atmosphere to suppress undesired phases.

In a brilliant recent review [35] titled: "Superconductivity at 100: where we've been and where we are going", Larbalestier and Canfield noted that: "*A special characteristic of the search for new materials is that it requires a certain kind of passionate optimism because there is no recipe for finding new materials that may manifest new types of superconductivity*". While we agree with the first part of this statement, we strongly disagree with the second part. Our recent papers [17 - 27] on the search for new superconductors indicate that there is a material specific solution to the prediction of new superconducting materials and their Tcs. Those results and the results in this paper stand to challenge and refute the "no-recipe-for-finding-new-materials" belief. Though serendipity has played a very significant role in past superconductivity discoveries [29, 30], we have entered the age of **materials by design [31] and by combinatorial material synthesis [36 - 39]. This paper and previous ones [24 -27] have shown the path to novel higher-Tc superconductivity via computational combinatorial synthesis. The search for new superconductors will be accelerated when computational and experimental material scientists collaborate in the quest for "the ultimate materials challenge": room temperature superconductivity [30, 32, 33] and a comprehensive material specific theory of superconductivity [34].**

## Conclusions

**Computational combinatorial synthesis of $F_Z$=4.67 and $F_Z$=12.67 may result in $F_Z$=8.67 family of compounds with Ne=2.67 and with 0.8< $Ne/\sqrt{Z}$ <1.0, suggesting superconductivity. These 4, 5 and 6-atom compounds with Fw/Z of approximately 12.4 strongly suggest high-Tc superconductivity [17]. Estimates indicate that some $F_Z$ = 8.67 family members may have Tcs as high as 60K. This newly discovered $F_Z$ = 8.67 family of over 160 materials provides further computational and experimental challenges of verifying their feasibility, stability and true Tcs.**

## Acknowledgements

**The author acknowledges stimulating and inspiring discussions with A.O.E Animalu on symmetry in mathematics, physics and chemistry and encouragement from M. Brian Maple to "continue the search". M.J. Schaffer sponsored this project while J.R. O'Brien provided literature links and relaxing discussions after culinary outings.**

| Material | | $\chi$ | Ne | Z | Ne/$\sqrt{Z}$ | Fw | Fw/Z |
|---|---|---|---|---|---|---|---|
| 1 | LiBC | 1.8333 | 2.6667 | 4.6667 | 1.2344 | 29.76 | 6.377 |
| 2 | $BeB_2$ | 1.8333 | 2.6667 | 4.6667 | 1.2344 | 30.63 | 6.564 |
| 3 | $Be_2C$ | 1.8333 | 2.6667 | 4.6667 | 1.2344 | 30.03 | 6.435 |
| 4 | $Li_2O$ | 1.8333 | 2.6667 | 4.6667 | 1.2344 | 29.88 | 6.403 |
| 5 | LiBeN | 1.8333 | 2.6667 | 4.6667 | 1.2344 | 29.96 | 6.420 |
| 6 | $LiB_5$ | 1.8333 | 2.6667 | 4.6667 | 1.2344 | 60.99 | 13.07 |
| 7 | $Li_3BN_2$ | 1.8333 | 2.6667 | 4.6667 | 1.2344 | 59.65 | 12.78 |
| 8 | $NaAlH_4$ | 1.8 | 1.3333 | 4.6667 | 0.6172 | 54.01 | 11.574 |
| 9 | $KBH_4$ | 1.8667 | 1.3333 | 4.6667 | 0.6172 | 53.95 | 11.56 |
| 10 | $MgH_2$ | 1.8 | 1.3333 | 4.6667 | 0.6172 | 26.33 | 5.642 |

**Table 1:** 10 MSCDs of some Z =4.67 materials with Ne=2.67 and 1.33

| | Material | $\mathcal{X}$ | Ne | Z | Ne/$\sqrt{Z}$ | Fw | Fw/Z |
|---|---|---|---|---|---|---|---|
| 1 | CaBeSi | 1.433 | 2.667 | 12.667 | 0.7493 | 77.18 | 6.09 |
| 2 | $Na_2S$ | 1.433 | 2.667 | 12.667 | 0.7493 | 78.05 | 6.16 |
| 3 | LiKS | 1.433 | 2.667 | 12.667 | 0.7493 | 78.11 | 6.17 |
| 4 | NaAlSi | 1.4 | 2.667 | 12.667 | 0.7493 | 78.06 | 6.16 |
| 5 | $Mg_2Si$ | 1.4 | 2.667 | 12.667 | 0.7493 | 76.71 | 6.06 |
| 6 | NaMgP | 1.4 | 2.667 | 12.667 | 0.7493 | 78.27 | 6.18 |
| 7 | KAlC | 1.6 | 2.667 | 12.667 | 0.7493 | 78.09 | 6.16 |
| 8 | NaCaN | 1.633 | 2.667 | 12.667 | 0.7493 | 77.08 | 6.09 |
| 9 | $LiBGe_{0.89}Si_{0.11}$ | 1.6 | 2.667 | 12.667 | 0.7493 | 85.46 | 6.75 |
| 10 | KBSi | 1.533 | 2.667 | 12.667 | 0.7493 | 78.0 | 6.16 |
| 11 | KBeP | 1.467 | 2.667 | 12.667 | 0.7493 | 79.09 | 6.24 |
| 12 | $MgAl_2$ | 1.4 | 2.667 | 12.667 | 0.7493 | 78.27 | 6.18 |
| 13 | LiAlTi | 1.333 | 2.667 | 12.667 | 0.7493 | 81.8 | 6.46 |
| 14 | LiCaP | 1.367 | 2.667 | 12.667 | 0.7493 | 77.99 | 6.16 |
| 15 | $Na_3AlP_2$ | 1.4 | 2.667 | 12.667 | 0.7493 | 157.89 | 12.465 |
| 16 | $K_3BN_2$ | 1.7333 | 2.667 | 12.667 | 0.7493 | 156.13 | 12.326 |

**Table 2:** 16 MSCDs of Z =12.667 materials.

| | Fz=4.67 materials | Fz=12.67 materials | Some 10 out of 160 Fz=8.67 possible materials |
|---|---|---|---|
| 1 | LiBC | CaBeSi | LiBeCaBCSi |
| 2 | $BeB_2$ | $Na_2S$ | $LiNa_2BCS$ |
| 3 | $Be_2C$ | LiKS | $Li_2KBCS$ |
| 4 | $Li_2O$ | NaAlSi | $Li_2Na_2OS$ |
| 5 | LiBeN | $Mg_2Si$ | $Li_2CaBeSiO$ |
| 6 | $LiB_5$ | NaMgP | $Li_3KOS$ |
| 7 | $Li_3BN_2$ | KAlC | $Li_2Mg_2SiO$ |
| 8 | $NaAlH_4$ | NaCaN | $Li_2NaCaNO$ |
| 9 | $KBH_4$ | $LiBGe_{0.89}Si_{0.11}$ | $Be_2Mg_2SiC$ |
| 10 | $MgH_2$ | KBSi | $Be_2NaCaNC$ |
| 11 | | KBeP | |
| 12 | | $MgAl_2$ | |
| 13 | | LiAlTi | |
| 14 | | LiCaP | |
| 15 | | $Na_3AlP_2$ | |
| 16 | | $K_3BN_2$ | |

**Table 3:** combinatorial synthesis of 10 Fz=4.67 with 16 Fz=12.67 materials may result in 160 new Fz=8.67 materials, a few of which are shown in this table.

| Fz=8.67 materials | | $\chi$ | Ne | Z | $Ne/\sqrt{Z}$ | Fw | Fw/Z | Tc (estimated with n=2.5) |
|---|---|---|---|---|---|---|---|---|
| 1 | LiBeCaBSiC | 1.633 | 2.667 | 8.667 | 0.9058 | 106.94 | 12.34 | 45.6 |
| 2 | $LiNa_2BCS$ | 1.633 | 2.667 | 8.667 | 0.9058 | 107.81 | 12.44 | 46.0 |
| 3 | $Li_2KBCS$ | 1.633 | 2.667 | 8.667 | 0.9058 | 107.87 | 12.45 | 46.0 |
| 4 | $Li_2Na_2OS$ | 1.633 | 2.667 | 8.667 | 0.9058 | 107.93 | 12.45 | 46.0 |
| 5 | $Li_2CaBeSiO$ | 1.633 | 2.667 | 8.667 | 0.9058 | 107.06 | 12.35 | 45.7 |
| 6 | $Li_2Mg_2SiO$ | 1.617 | 2.667 | 8.667 | 0.9058 | 106.59 | 12.30 | 45.7 |
| 7 | $Na_2Be_2CS$ | 1.633 | 2.667 | 8.667 | 0.9058 | 108.08 | 12.47 | 46.0 |
| 8 | $Be_2Mg_2SiC$ | 1.617 | 2.667 | 8.667 | 0.9058 | 106.74 | 12.32 | 45.1 |
| 9 | $Li_2NaCaNO$ | 1.733 | 2.667 | 8.667 | 0.9058 | 106.96 | 12.34 | 48.5 |
| 10 | $Be_2NaCaNC$ | 1.733 | 2.667 | 8.667 | 0.9058 | 107.11 | 12.36 | 48.5 |

**Table 4:** 10 MSCDs of some Z =8.67 materials with Ne=2.67 out of 160 possibilities. The Tc can be estimated with the Formula: $T_c = \chi \frac{Ne}{\sqrt{Z}} K_o$ derived in reference [17]. Ko =n(Fw/Z). n=2.5 in the above calculations. For MgB$^2$, n=3.65. The actual Tcs may be higher!

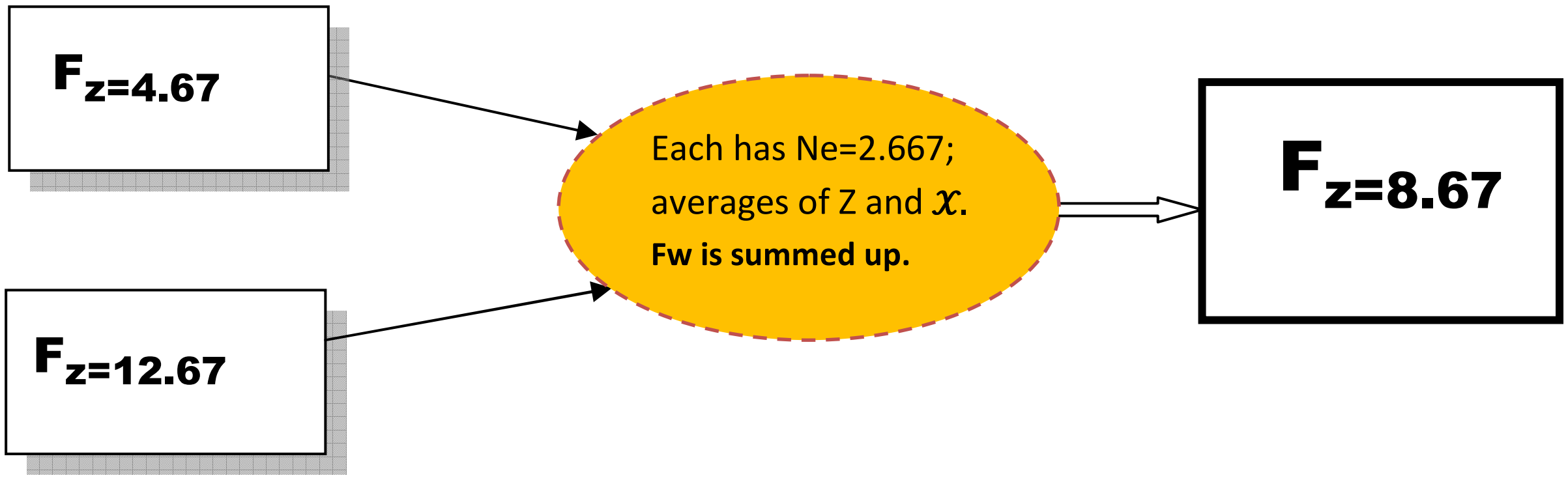

**Figure 1**: Combining materials from Fz=4.67 and Fz=12.67 may yield a new family Fz=8.67 material with Ne=2.67 and Z=8.67. The final electronegativity will be an average of the electronegativities of the Fz=4.67 and Fz=12.67 materials.